\newcommand{\PaperTitle}{Topic-wise Exploration of the Telegram Group-verse}

\documentclass[sigconf]{acmart}

\usepackage{subcaption}
\usepackage{footmisc}
\usepackage{lscape}
\usepackage{float}
\usepackage{rotating}
\usepackage{geometry}
\usepackage{tablefootnote}
\usepackage[normalem]{ulem}

\copyrightyear{2025}
\acmYear{2025}
\setcopyright{cc}
\setcctype{by}
\acmConference[WWW Companion '25]{Companion Proceedings of the ACM Web
Conference 2025}{April 28-May 2, 2025}{Sydney, NSW, Australia}
\acmBooktitle{Companion Proceedings of the ACM Web Conference 2025 (WWW
Companion '25), April 28-May 2, 2025, Sydney, NSW, Australia}
\acmDOI{10.1145/3701716.3717506}
\acmISBN{979-8-4007-1331-6/2025/04}

\begin{document}

\title{\PaperTitle}

\author{Alessandro Perlo}
\affiliation{%
  \institution{Politecnico di Torino}
  \city{Torino}
  \country{Italy}
}
\email{alessandro.perlo@studenti.polito.it}

\author{Giordano Paoletti}
\affiliation{%
  \institution{Politecnico di Torino}
  \city{Torino}
  \country{Italy}
}
\email{giordano.paoletti@polito.it}

\author{Nikhil Jha}
\affiliation{%
  \institution{Politecnico di Torino}
  \city{Torino}
  \country{Italy}
}
\email{nikhil.jha@polito.it}

\author{Luca Vassio}
\affiliation{%
  \institution{Politecnico di Torino}
  \city{Torino}
  \country{Italy}
}
\email{luca.vassio@polito.it}

\author{Jussara Almeida}
\affiliation{%
  \institution{Universidade Federal de Minas Gerais}
  \city{Belo Horizonte}
  \state{Minas Gerais}
  \country{Brazil}
}
\email{jussara@dcc.ufmg.br}

\author{Marco Mellia}
\affiliation{%
  \institution{Politecnico di Torino}
  \city{Torino}
  \country{Italy}
}
\email{marco.mellia@polito.it}

\renewcommand{\shortauthors}{Alessandro Perlo et al.}

\begin{abstract}
Although Telegram is currently one of the most popular instant messaging apps in the world, previous studies have mainly focused on analysing discussions on specific angles and topics. In this paper, we present a broad analysis of publicly accessible groups that cover a wide range of discussions, including Education, Erotic, Politics, and Cryptocurrencies. How do people interact with different topic groups? Is there any common or peculiar behaviour? We engineer and offer an open-source tool to automate the collection of messages from Telegram groups, a non-straightforward problem. We use it to collect more than 51 million messages from 669 groups. Here, we present a first-of-its-kind, per-topic analysis, contrasting the users' activity patterns from different angles --- the language, the presence of bots, the type and volume of shared media content, links to external platforms, etc. Our results confirm some anecdotal evidence, e.g., indications of spamming behaviour, and unveil some unexpected findings, e.g., the different sharing patterns of video and message length in groups of different topics. Our research provides a horizontal analysis of the public group in Telegram across various general topics, establishing a foundation for future studies that can delve deeper into user interactions and content dynamics within this unique messaging environment.
\end{abstract}



\maketitle

\section{Introduction}

Telegram has experienced remarkable growth in the past years, becoming one of the most popular instant messaging apps in the world. In July 2024, it surpassed the mark of 950 million monthly active users worldwide\footnote{Telegram CEO Pavel Durov https://t.me/durov/337}. Telegram offers several features to its users, who can organize themselves into different spaces of communication such as private chats (one-to-one), groups (many-to-many) or channels (one-to-many). 

Yet, the literature on Telegram is still limited in breadth. Targeting publicly accessible groups and channels, most prior works focused on {\it textual} content (e.g. news, hate speech), specific groups (e.g., terrorists~\cite{yayla2017telegram}) or countries (e.g., Iran \cite{hashemi_telegram_2019}), and a single topic of discussion (e.g., far-right politics~\cite{Urman_Katz_2022}).

In contrast, 
this paper is driven by the following research questions: (i) how do users behave on Telegram groups in terms of the features they most often make use of to interact with others (e.g., video sharing, link sharing, use of reactions, polls, etc.)? (ii) How do such platform usage patterns change across groups discussing different topics?

We develop an open-source crawler designed to access public Telegram groups, collect historical messages up to a specified date, and continuously update the message collection over time, to build a longitudinal archive.
We use the crawler to gather data from more than a thousand open groups and focus on those with at least 100 active users, distributed across 10 different topics of discussion including {\it Politics}, {\it Cryptocurrency}, \textit{Darknet}, \textit{Erotic}, {\it Video and Films}, etc. 
In total our data covers around 51.6~M messages and 1.4~M distinct users over a two-month observation period.

We analyse our data aiming to measure and contrast user activity patterns in groups across different topics.
We analyse the mix of various languages, the footprint of official Telegram bots, the diverse habits in sharing media (e.g., videos, audios, images, GIFs) and external content via URLs. We overall witness peculiar behavioural patterns; some might be expected (e.g., the presence of automated user behaviour to programmatically share context) while others are more surprising (e.g., users in \textit{Darknet} post much longer messages than users in the other topics).

This paper is structured as follows. Section~\ref{sec:related} lists related work. in Section \ref{sec:method}, we present the architecture of our crawler
. 
In Section \ref{sec:results}, we detail the data collection process and present a first-of-its-kind characterisation of platform usage across different discussion topics. 
Section \ref{sec:media} further elaborates on the analysis of video content and URLs sharing patterns, together with a deeper analysis on the time elapsed between a YouTube video publication and its first appearance in Telegram groups under observation. 
Last, we draw conclusions in Section~\ref{sec:conclusion} and discuss ethics of the work in Section~\ref{sec:ethics}.

All in all, this work shows the heterogeneity of usages people exhibit on Telegram, corroborating known facts and exposing surprising findings. We hope this work stimulates other works in exploring some of the highlighted results. For this, we make both the data and the crawler open to the community.\footnote{\label{code} The code and data are available at \url{https://anonymous.4open.science/r/TopicWiseTelegram-7A81}}
\section{Related work}
\label{sec:related}

The characteristics and dynamics of messaging platforms have attracted a lot of attention. Notably, prior studies analysed content properties \cite{resende:2019:WWW, resende:2019:WebSci,maros_analyzing_2020} and information spread \cite{caetano:2019,nobre:2021} on WhatsApp's groups, hinting at the catalytic role of the platform in various real-world events \cite{garimella:2018,Bowles:2020,machado:2019}.

More recently, attention has been dedicated to groups and channels on Telegram, as the platform's popularity increases across the globe. Some studies were interested in the inner workings of the mobile application~\cite{satrya_digital_2016, anglano_forensic_2017}, and its use by particular user populations, such as  Iranian immigrants \cite{nikkhah_telegram_2018}, terrorist organizations \cite{yayla2017telegram,Alrhmoun_Winter_Kertész_2024}, extremist groups \cite{Hoseini_Melo_Benevenuto_Feldmann_Zannettou_2023,Kloo_Cruickshank_Carley_2024},  or particular countries (e.g., Iran and Russia \cite{hashemi_telegram_2019,akbari_platform_2019,Hanley_Durumeric_2024}). Others analysed the formation of communities within Telegram channels \cite{Urman_Katz_2022,tikhomirova_community_2021} and their connection to information spread \cite{venancio24, Hoseini_Melo_Benevenuto_Feldmann_Zannettou_2024, dargahi_nobari_characteristics_2021}.  Some other efforts studied content properties, limiting to textual content and usage patterns, with attention given to news content \cite{naseri_analyzing_2019}, hate speech and abusive language~\cite{wich_introducing_2022}, as well as the presence of fake channels (i.e., those impersonating important services or persons) \cite{morgia_its_2023}. The use of Telegram to perform illicit activities (e.g., pump-and-dump activities in cryptocurrency markets \cite{xu_anatomy_2019}, manipulation of social media popularity \cite{weerasinghe_pod_2020}) has also been previously addressed. Overall, previous works focused on the information people exchange on Telegram, in groups and channels of a specific topic.   
Only Morgia \textit{et al.}~\cite{morgia_its_2023} used TGStat to gather channels associated with multiple topics. Yet, they did not distinguish between such topics and aggregated all of them to discover fake channels. 

In contrast, we here offer a topic-wise analysis of common user activities in Telegram groups. Rather than focusing on the type of information exchanged, or how it spreads, we show how people leverage different features (e.g.,  media types, links to external sites) to interact with each other, and how such patterns differ depending on the topics and goals of group discussion.

\section{Crawler and data collection}
\label{sec:method}

Given our interest in {\it user} behaviour, we focus our data collection effort on Telegram \textit{public groups}, i.e., public chats where all the members can send messages. 


To collect the data, we design an open-source, two-stage crawler that we offer to the community.\footref{code} 
At the first stage, the tool periodically crawls the TGStat website to discover public Telegram groups on various topics. At the second stage, the tool crawls Telegram by joining the discovered groups and collecting all messages.


\subsection{TGStat crawling}

TGStat is a service that catalogues popular Telegram groups and channels worldwide. 
Currently, TGStat's database covers almost 1.9~M channels and groups \cite{tgstatstat}, which are categorised into 48 pre-defined topics.
For each topic, TGStat shows the lists of the top-100 groups according to various metrics. These lists are continuously and dynamically updated. 
TGStat characterises each group by some metadata, including the group name, topic, language, and the monthly Active Users (AU), i.e., the number of unique users who have shared messages in the group in the past month.



We extract information from TGStat engineering a Python-based crawler using the BeautifulSoup package~\cite{beautifulsoup}. We periodically run it to automatically extract the lists of groups in various topics.
This allows us to grow the group lists in those topics of interest to us (see discussion in Section~\ref{subsec:crawling}).

Most prior studies of Telegram searched for links to existing groups in social media, news and even word-of-mouth~\cite{Hoseini_Melo_Benevenuto_Feldmann_Zannettou_2023,baumgartner_pushshift_2020,dargahi_nobari_characteristics_2021}. 
This approach can demand extensive crawling, particularly for diverse topics. TGStat streamlines this process by offering categorized groups with real-time activity metrics, facilitating targeted selection and per-topic analysis. Prior studies~\cite{tikhomirova_community_2021, morgia_its_2023, Urman_Katz_2022} leveraged TGStat but relied on static snapshots. In contrast, we continuously expand our dataset over multiple days. Furthermore, given our focus on per-topic analysis, we assess the reliability of TGStat’s categorization—a crucial step overlooked in past work (see Section~\ref{sec:topic-verification}).

\subsection{Telegram crawling}

Given a list of previously discovered groups, our crawler automatises the group join and message collection tasks. We rely on the Telethon Python package~\cite{telethon} and design a scalable tool based on threads: a master instructs workers to \textit{join} (and leave, if desired) a group, \textit{check} if a pending request for join has been accepted, \textit{collect} new messages, or just \textit{wait}. For scalability, we use multiple Telegram IDs, each associated with a worker.
The master keeps a list of groups to collect messages from and instructs workers to do so from a desired initial date until the present. We store the collected information in a MongoDB database for later processing.

We instrument our crawler to join and stay in groups as we discover them on TGStat. To refresh the collection of messages, workers download only the new messages since the last retrieved snapshot. 
For every group and message, the crawler stores all the returned information in JSON format in the MongoDB instance. In this paper, we focus on the following message information: sender user's identifier, message body, message time, and media contained in the message (image, video, GIF, poll, etc.).


\subsection{Crawler design challenges}
Telegram implements several countermeasures to avoid API abuse, notably:  i) a limit of 500 groups a given Telegram ID can join; ii) an unspecified upper limit on the rate to join new groups which, if not respected, 
causes a lengthy temporary ban \cite{telegram-error}; 
iii) a without-any-notice permanent ban of novel-activated Telegram IDs that start to interact with the platform with high frequency. 
Respecting these limitations requires ingenuity when designing the crawler. First, we declared our intentions to the official Telegram support channel. 
Second, we carefully controlled the group joining rate to limit the temporal ban. Third, we used multiple already-active Telegram IDs, each associated with a worker thread to scale the data gathering.

Telegram offers the possibility of setting up administration bots (known as \textit{Telegram bots}) that ease group management. Captcha protection bots are popular for filtering fake user bots, i.e., actual Telegram accounts used to programmatically spam messages in open groups. Such captcha protection bots may kick users out if they do not solve the captcha after a specific time. Other bots or administrators might enforce different rules or criteria for group participation. Whenever we were removed from a group, we respected the administrators' willingness and did not try to join the group again. 
Similarly, to respect the privacy indications of the group administrators, we only consider groups where the participants' messages are persistent and not automatically removed (``auto-delete'' functionality removes messages 24 hours or 7 days after sending).


Our crawler can collect up to thousands of messages per second per worker and join tens of groups per hour without overcoming Telegram rate limitations.

\section{Topic Characterisation}
\label{sec:results}


\subsection{Data collection and filtering}  \label{subsec:crawling}

\subsubsection{Topic selection}
On {April 1\textsuperscript{st}, 2024} we collected the top-100 groups for all TGStat 48 topics. Out of these, we select the 12 topics in which we were able to join at least 10 English language groups, a condition that allows us to manually validate the accuracy of TGStat's topic labelling.
We keep crawling TGStat every week to refresh the lists of top-100 groups for these 12 topics to observe how those lists change over time. We stop on May 1\textsuperscript{st}, 2024, discovering 1,368 groups in total. The growth in the number of groups discovered over time is notable in specific topics such as \textit{Erotics}, \textit{Cryptocurrencies} and \textit{Bookmaking}, where we find around 20\% new groups every week. This illustrates that taking a single snapshot from TGStat, as prior work~\cite{tikhomirova_community_2021, morgia_its_2023,Urman_Katz_2022}, would limit the lists of discovered groups.

Feeding these growing lists to the Telegram crawler, we find that 8.6\% of groups have the auto-delete function enabled. We abandon them immediately. We also fail to join 18.8\% of groups because the group (i) did not exist anymore, (ii) changed their name before we could join it, or (iii) are moderated and either the administrator did not admit us or a bot kicked us out after joining. We thus successfully join and monitor 993 groups. For each tracked group, we collect all messages starting from {March 1\textsuperscript{st}} to {April 30\textsuperscript{th}}.\footnote{For this work, we limit the collection period to avoid overloading the Telegram servers.} In total, we collect over 50~M messages, with about 1~M new messages gathered each day.

\begin{table}
    \centering
    \footnotesize
    \caption{Dataset statistics for verified topics. Column 2 refers to groups discovered in TGStat, all others refer to joined groups with at least 100 Active Users (AU) with consistent topic.}
    \begin{tabular}{l|c c|r r|r r}
    \toprule
        \multicolumn{1}{c}{\textbf{Topic}} & 
        \multicolumn{2}{c}{\textbf{\# Groups}} & 
        \multicolumn{2}{c}{\textbf{Average $\#$ Users}} & 
        \multicolumn{2}{c}{\textbf{\textbf{$\#$ Messages}}} \\
          & 
         TGStat & 
         \multicolumn{1}{p{0.8cm}|}{\centering Joined} & 
         \multicolumn{1}{p{0.8cm}}{\centering Memb.\\(k)} &
         \multicolumn{1}{p{0.6cm}|}{\centering AU \\ (\%)} &
         \multicolumn{1}{p{0.6cm}}{\centering Tot. \\ (M)} &
         \multicolumn{1}{p{0.6cm}}{\centering Per\\ AU} \\
        \midrule
        \textit{Education}  & 115 & 98 & 28.7 & 8.5 & 4.8& 21.1\\
        \textit{Bookmaking} & 120 & 91 & 16.6 & 13.1 & 9.0& 47.2\\
        \textit{Crypto} & 123 & 80 & 69.6 & 9.0 & 10.3& 22.6\\
        \textit{Technologies} & 108 & 67 & 27.0 & 7.8 & 6.0 & 43.0\\
        \textit{Darknet} & 112 & 62 & 11.0 & 19.2 & 5.2& 41.9\\
        \textit{Software} & 114 & 61 & 14.7 & 7.7 & 3.8 & 55.8\\
        \textit{Video\&films} & 114 & 59 & 13.1 & 9.2 & 2.4 & 34.8\\
        \textit{Politics} & 115 & 58 & 4.9 & 26.0 & 4.2& 57.1\\
        \textit{Erotic} & 124 & 52 & 22.5 & 7.7 & 4.6& 56.1\\
        \textit{Linguistics} & 109 & 40 & 6.7 & 8.0 & 1.4 & 67.3\\
        \midrule
        Total & 1,154\tablefootnote{The total number of groups adds up to 1,368 if we consider 214 groups from \textit{Courses and guides} and \textit{Economics} that we initially considered and later discarded.} & 669 & 23.7 & 9.7 &  51.7 & 36.4\\
    \bottomrule
    \end{tabular}
    \label{tab:Topic}
\end{table}

\subsubsection{TGStat topic verification}
\label{sec:topic-verification}

Next, we verify if the per-topic categorisation provided by TGstat is reliable. To that end, we check if the {\it actual}  topic of discussion is (i) coherent with the topic assigned by TGStat and (ii) consistent across time. We pick all English groups (206) and for each group, we select three sets of 30 consecutive messages, each set separated by the others by ten days. The 206 groups where randomly assigned to three human evaluators, each required to independently assess whether each set of messages was, collectively, addressing a subject that was consistent with the topic of the group. To measure the agreement among evaluators, we use the Fleiss' Kappa \cite{fleiss1981measurement}. The labelling showed an agreement of 0.71, indicating substantial agreement between evaluators.

TGStat's topic assignment proves mostly correct, with two exceptions: the \textit{Courses and guides} groups are mostly filled with spam; and the \textit{Economics} groups mostly host discussions about cryptocurrencies, for which a dedicated topic already exists. We thus discard the 166 groups in these two topics, ending up with 827 groups.

\subsubsection{Selecting active groups}
To guarantee groups are active and with enough diversity, we keep groups that have at least 100 active users, i.e. users who sent at least one message in the two-month observation period.
From the 827 groups, we discard 158 of them, ending with 669 groups, as detailed in column 3 of Table~\ref{tab:Topic}. For comparison, the total number of groups discovered in TGStat in each topic is shown in column 2.
The table also details the average number of users per group, the percentage of active ones, the total number of messages and the average number of messages per active user.  Figures vary widely, showing already very different interests, engagement and activity levels across the topics.

For the sake of completeness, Figure~\ref{fig:crawler_result} in the Appendix~\ref{sec:app} details the breakdown of the various cases 
we faced when trying to collect messages from Telegram. Depending on the topic, we observe various failure cases which may significantly reduce the number of groups to follow.

\subsection{Per-Topic Characterisation}
We characterize our dataset by extracting various features from each group and then aggregating them on a per-topic basis (i.e., per-group macro average). This allows us to avoid the bias induced by large groups. 
Our goal is to explore how differently users interact on each topic.

\subsubsection{Telegram Bot usage and user activity}
\label{subsec:bot}

\begin{figure}
    \centering
    \includegraphics[width=.65\columnwidth]{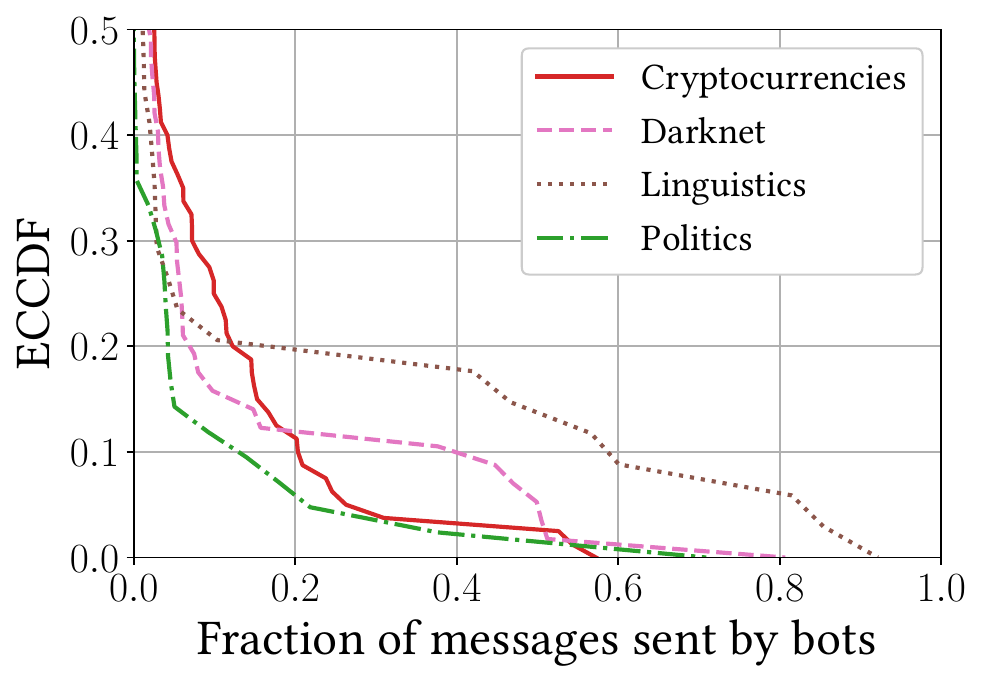}
    \caption{Fraction of messages sent by bots in a group across topics.}
    \label{fig:bot-ccdf}
\end{figure}

Telegram group administrators can incorporate bots\footnote{Note that we use the term \textit{bot} to refer exclusively to official bots and not to users exhibiting bot-like automated behaviour.} into their groups, offering a wide set of functionalities, from welcoming newcomers with group rules to responding to user commands, from collecting statistics to moderating join requests and messages.
Telegram bots are quite popular: only 10.2\% of groups in our dataset do not include bots; in the median, there are four bots per group. Interestingly, 1.5\% of groups have 20 bots (the maximum allowed by Telegram). Some bots enjoy significant popularity: \textit{Combot} \cite{combot} and \textit{MissRose\_bot} \cite{missrose} are present in 145 and 129 groups, respectively. Both provide moderation services, analytics, and anti-spam features.


Bots footprint is not negligible: they generate on average 8.6\% of messages, with notable variations across topics. Figure~\ref{fig:bot-ccdf} shows, for 4 topics, the Empirical Complementary Cumulative Distribution Function (ECCDF) of the fraction of messages sent by bots for different groups. 
The largest fractions of messages sent by bots are in \textit{Linguistics} groups.
Some of these bots are integral to learning platforms and merit examinations. For instance \textit{Quizbot} \cite{quizbot} is widely deployed in \textit{Linguistics} and \textit{Education} groups (27.5\% and 28.6\%, respectively). It generates 39.0\% and 13.0\% of messages. In one \textit{Linguistics} group it generates 91\% of messages. Conversely, \textit{Politics} groups see the smallest fraction of messages generated by bots (green dotted line), possibly testifying to a higher user engagement\footnote{\label{note1}For messages sent by a user account, we cannot distinguish between messages sent by a human or by an automated system.} in political groups than in other topics. 
Curiously, there is a quite large fraction of groups with bots that simply collect statistics or moderate the group without sending any message (leftmost part of Figure~\ref{fig:bot-ccdf}).  
For the remainder of our analysis, we remove messages sent by Telegram bots.

Focusing on the number of messages actual users generate, we observe that there are a few users sending thousands of messages, while the majority are not active (see column 4 in Table~\ref{tab:Topic}) or send few messages. Indeed, the Empirical Probability Density Function (EPDF) of the number of messages generated by users follows a heavy-tailed shape that can be fit by a Pareto distribution with $\alpha=1.9$.  Remarkably, the fittings of the per-topic EPDFs are very similar, suggesting the universality of these behaviours, as widely acknowledged in the literature\cite{barabasi2005origin,rybski2009scaling}.

\textit{\textbf{Main takeaways:} official bots' impact on the activity of Telegram is significant (8.6\% of messages), with some moderation bots being present on a wide variety of groups. Some topics are more prone to bots offering also specific features (e.g., \textup{QuizBot} generating 39\% of messages in \textup{Linguistics}' groups}.

\subsubsection{Language}

\begin{figure}
    \centering
    \includegraphics[width=.6\columnwidth]{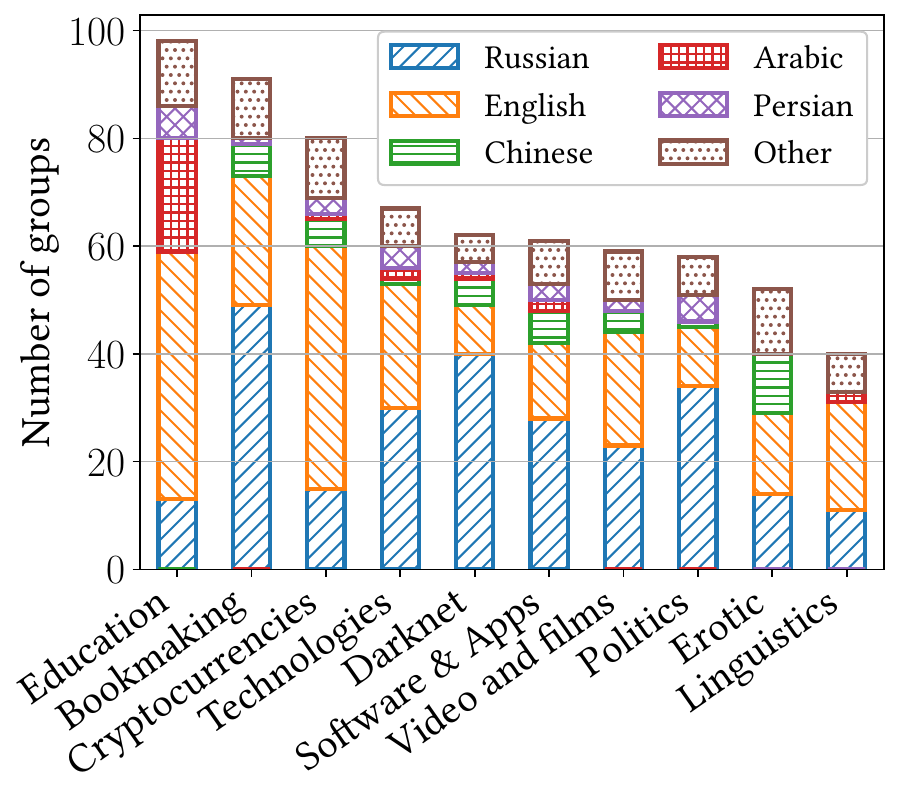}
    \caption{Most popular language in each group.  }
    \label{fig:language-composition}
\end{figure}
The next question we answer is what are the languages people use in each topic and group. For each {\it textual} message in a group, we associate a language by employing the FastText language identification library~\cite{joulin2017bag} to obtain the distribution of languages for each group.
In Figure~\ref{fig:language-composition} we report the breakdown of the {\it most popular} language for groups on the same topic:

$\bullet$ English (a global language) and Russian (being Telegram very popular in Russia) are the two most popular languages.
Their share changes based on the topic. For instance, most groups in \textit{Bookmaking} and \textit{Darknet} have a lot of Russian groups. Conversely, the majority of \textit{Education} and \textit{Cryptocurrency} groups are in English, possibly due to the worldwide interest in such topics.


 $\bullet$ Despite Telegram's restricted use in Iran, some popular groups are in Persian, mostly in \textit{Politics}. This is in line with the claim that Telegram is used 
 to evade state censorship in Iran~\cite{akbari_platform_2019}.

 $\bullet$ Although Telegram is blocked in China \cite{telegram-china}, 
we do find groups with Chinese as the dominant language (up to 21\% of \textit{Erotic} groups).

We observe that in almost half of the groups, more than 75\% of the messages are written in the same dominant language. Still, 
we observe messages written in other languages, hinting at a global user population. 
\textit{Linguistics} is the topic where groups contain the largest mix of languages, which supports the anecdotal observation of users practising foreign languages and mixing messages in their native language in such groups. In contrast, both \textit{Darknet} and \textit{Technology} stand out with more than 58\% of the groups having more than 80\% of the messages in a single language. 

\textit{\textbf{Main takeaways:} English and Russian are the most widely used languages, dominant in 34.1\% and 38.4\% of groups, respectively. Notably, Persian (3.9\%) and Chinese (6.0\%) also appear frequently, despite Telegram restrictions in Iran and China. Many groups mix multiple languages within their discussions.}

\subsubsection{Message length}

\begin{figure}[t]
    \centering
    \includegraphics[width=0.65\columnwidth]{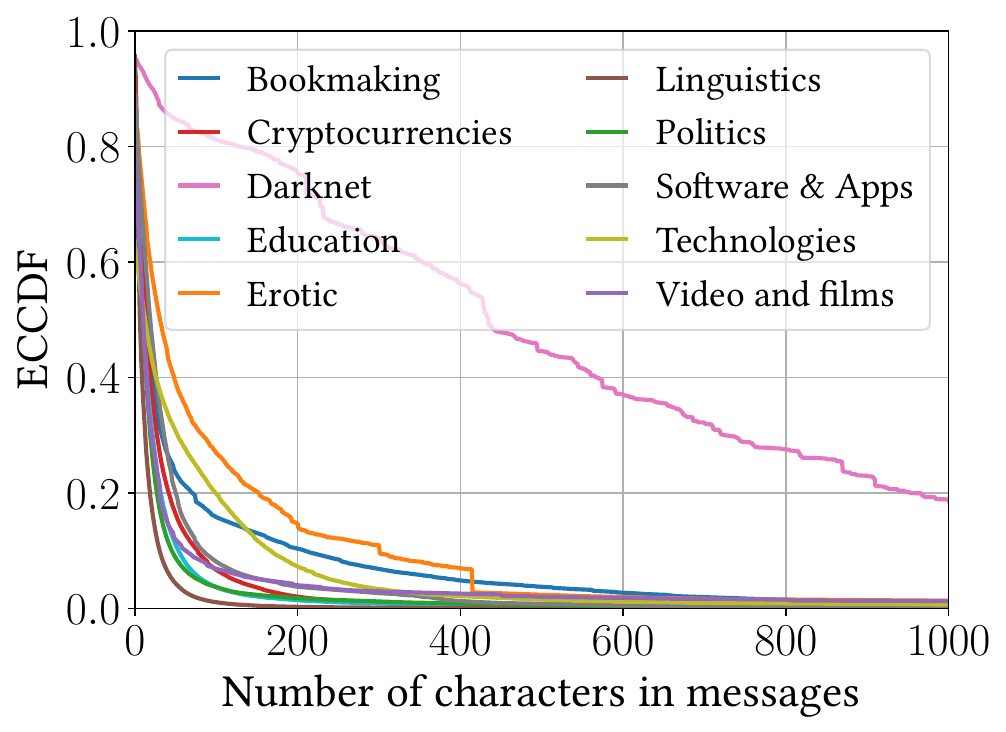}
    \caption{ECCDF of the number of characters in text messages grouped by topic (groups in English).}
    
    \label{fig:message-length-ccdf}
\end{figure}


Figure~\ref{fig:message-length-ccdf} shows the ECCDF of the length of {\it textual} messages for each topic. Since message length may be influenced by language, we consider only messages written in English-dominated groups. We observe some great distinctions across topics. On one hand, \textit{Darknet} groups are dominated by very long messages (80\% longer than 100 characters). A manual check unveils that most messages contain samples of illicit content people trade, such as bank account credentials, credit cards, passports and SSNs. Conversely, in \textit{Bookmaking} and \textit{Technologies} groups, long messages describe bookmaking websites, betting experiences and results, or offer details about devices for sale. In \textit{Erotic} groups, long messages are used to advertise services and sell content, some of them repeated multiple times (observe the steps in the ECCDF).
On the other hand, \textit{Linguistics} and \textit{Politics} groups are dominated by very short messages in which people debate or provide quick suggestions (80\% shorter than 30 characters). 
At last, steps in the distribution might suggest the presence of repeated automated messages (mostly spam), further analysed in Section \ref{subsec:repeated_sharing}.

\textit{\textbf{Main takeaways:} 
There appears to be a correlation between message length and promotional content, with longer messages linked to advertisements, while discussion-driven groups to shorter messages (e.g., compare Bookmaking and Linguistics in Figure~\ref{fig:message-length-ccdf}).}


\subsubsection{Usage of non-textual elements}

\begin{figure}
    \centering
    \includegraphics[width=0.8\columnwidth]{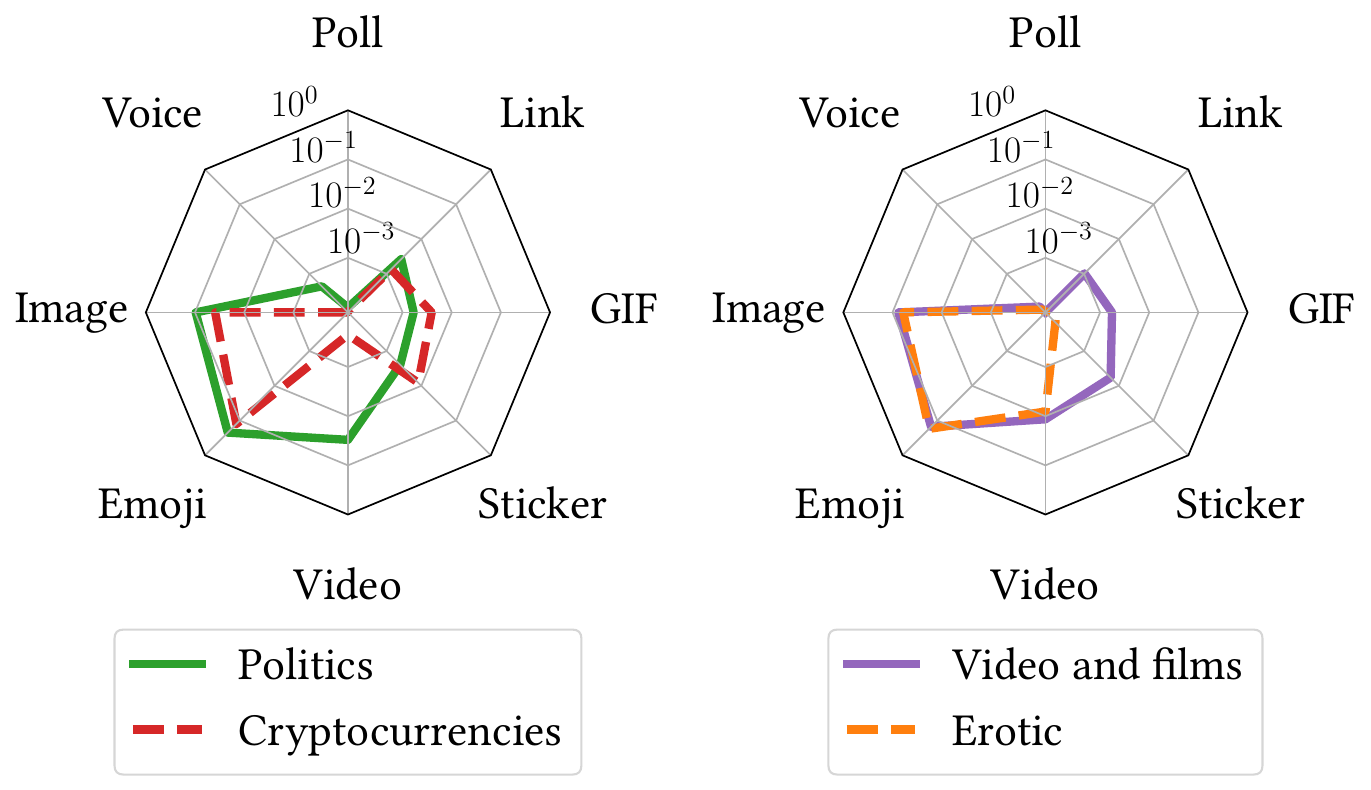}
    \caption{Median fraction of messages with non-textual elements in selected topics.
    }
    \label{fig:media-spiderplots}
\end{figure}

We now broaden our analysis to consider non-textual elements. 
Specifically,  we extract, for each group, the fraction of messages containing images, external links, voice messages, polls, GIFs, stickers, videos, and emojis. How are these elements used? To gauge this, we compute the fraction of messages with such element in a group and compute the average over all groups of a given topic (macro average). 
Figure~\ref{fig:media-spiderplots} visually compares two pairs of selected topics using radar charts. Table~\ref{tab:details} and Figure~\ref{fig:appendix-spiderplots} in Appendix~\ref{sec:app} provides the complete set of results. Some interesting findings emerge:

 $\bullet$ {\it  Politics} groups represent the typical usage of non-textual elements: 20--30\% of messages contain emojis; 10\% of messages share an image; stickers are more popular than GIFs; few messages contain voice content; polls are mostly an unused feature (polls are in fact only used in {\it Linguistics} groups).

 $\bullet$ {\it  Cryptocurrencies}  groups represent some mixed usage: no videos and voice messages, fewer photos and emojis but more stickers and GIFs than average.

 $\bullet$ Groups in {\it Video and Films } and  {\it Politics} have very similar usage patterns (i.e. radar shape), though, surprisingly, the former has fewer videos (which are present in only 1\% of the messages) --- see Section~\ref{sec:media}.


 $\bullet$  
 {\it Erotic} groups present the minimum usage of non-textual elements: no stickers, no GIFs, while photos are found in 6\% of messages. 
 Surprisingly, we see very few links to external platforms. In fact, readers are invited to contact advertisers via private chat.

Notice that sending stickers requires manual actions hardly automatizable. 
Also, stickers are commonly used as reactions to other messages. Their prominent use in a group or topic may testify to a larger fraction of messages being sent by real users, or to a more confidential exchange: \textit{Erotic}, \textit{Darknet} and \textit{Technologies} have the least fraction of stickers and are dominated by ad-style messages. Conversely, in \textit{Bookmaking} and \textit{Cryptocurrencies} people use more stickers as reactions for suggestions. This argument was highlighted on other platforms such as WhatsApp where, for instance, in political groups, stickers are rarely forwarded, suggesting that users often save and maintain their collection of preferred stickers for future use rather than relying on the platform's sharing tools \cite{melo2024sticker}, further highlighting a more human-based interaction pattern.

\textit{\textbf{Main takeaways:} the media-sharing patterns across topics show big differences based on the topic under observation. No common habits emerge.}





\begin{table}
\centering
\footnotesize   
\caption{Per-topic detailed metrics of usage of non-textual elements in messages.}
\begin{tabular}{l|rrrrrrrr}
\toprule
\textbf{Topic} & 
\multicolumn{1}{p{0.3cm}}{\centering\textbf{Poll (\%)}} & 
\multicolumn{1}{p{0.4cm}}{\centering\textbf{Voice (\%)}} & 
\multicolumn{1}{p{0.4cm}}{\centering\textbf{Image (\%)}} & 
\multicolumn{1}{p{0.4cm}}{\centering\textbf{Emoji (\%)}} & 
\multicolumn{1}{p{0.4cm}}{\centering\textbf{Video (\%)}} & 
\multicolumn{1}{p{0.4cm}}{\centering\textbf{Sticker (\%)}} & 
\multicolumn{1}{p{0.4cm}}{\centering\textbf{GIF (\%)}} & 
\multicolumn{1}{p{0.4cm}}{\centering\textbf{Link (\%)}} \\ \hline
\textit{Education} & 0.01 & 0.02 & 5.13 & 12.43 & 0.04 & 0.01 & $<$0.01 & 0.06  \\
\textit{Bookmaking} & $<$0.01 & 0.03 & 8.67 & 19.81 & 0.40 & 0.72 & 0.38 & 0.04  \\
\textit{Crypto} &  $<$0.01 & $<$0.01 & 3.89 & 12.54 & 0.04 & 0.81 & 0.39 & 0.14  \\
\textit{Technologies} &  $<$0.01 & $<$0.01 & 6.12 & 8.22 & 0.10 & 0.09 & 0.05 & 0.29  \\
\textit{Darknet} &  $<$0.01 & $<$0.01 & 5.44 & 56.51 & 0.34 & $<$0.01 & 0.03 & 0.02 \\
\textit{Software} &  $<$0.01 & $<$0.01 & 6.50 & 9.61 & 0.42 & 0.42 & 0.10 & 0.36  \\
\textit{Video\&Films} & $<$0.01 & 0.02 & 7.61 & 15.10 & 1.15 & 0.58 & 0.17 & 0.10  \\
\textit{Politics} & $<$0.01 & 0.07 & 9.69 & 22.17 & 3.03 & 0.25 & 0.17 & 0.26 \\
\textit{Erotic} & $<$0.01 & $<$0.01 & 6.05 & 17.29 & 0.80 & 0.03 & 0.02 & $<$0.01  \\
\textit{Linguistics} & 0.10 & 0.08 & 2.34 & 15.34 & 0.33 & 0.16 & $<$0.01 & 0.04  \\ \hline
\textit{All Topics} & $<$0.01 & $<$0.01 & 6.02 & 14.65 & 0.23 & 0.23 & 0.09 & 0.07 \\ \bottomrule
\end{tabular}
\label{tab:details}
\end{table}

\vspace{-5pt}
\section{Multimedia and external links}
\label{sec:media}
We now delve deeper into the usage of specific non-textual elements, notably shared videos and URLs to external sites.

\subsection{Video size and duration}

We focus our analysis on the three topics with the largest share of videos: \textit{Politics}, \textit{Video and Films}, and \textit{Erotic}. Although  Figure \ref{fig:media-spiderplots} suggests some similarities in the amount of videos people directly share on Telegram, a closer examination of the video duration and file size reveals noteworthy differences in goals and types of shared videos, again showcasing notable differences in users' behavioural patterns among different topics.

\begin{figure}[t]
    \centering
    \includegraphics[width=0.8\columnwidth]{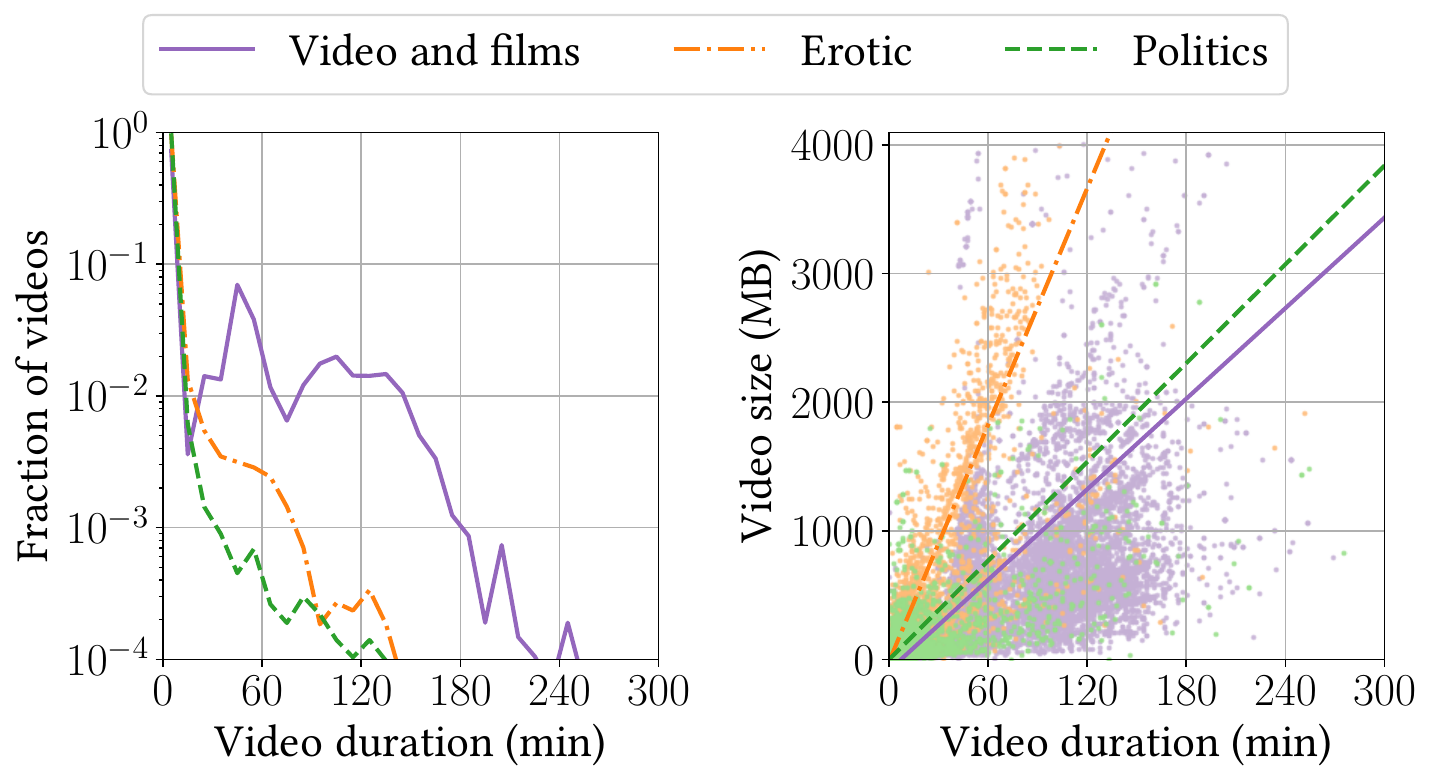}
    \caption{Distribution of the video duration (on the left, bin every 10 minutes) and comparison between video size and video duration (on the right, with regression line reported).}
    \label{fig:video-size-duration}
\end{figure}

The left plot of Figure~\ref{fig:video-size-duration} compares the video duration, by showing the fraction of videos by length. Observe how videos shared in {\it Video and Films} are notably longer than in other topics, with peaks roughly around the 60- and 120-minute marks (notice the log-y scale). Manual inspection confirms that people share entire TV series episodes and movies. Their total volume amounts to 7.45~TB of data. \textit{Erotic} and \textit{Politics} tend to have much shorter videos.

Compare now this video duration with its file size --- see the right graph of Figure~\ref{fig:video-size-duration}, a scatter plot with regression lines, where each point represents a shared video. Again, we observe a noticeable difference between topics: videos shared in {\it Erotic} groups have significantly higher video rate (thus quality), despite shorter durations, as evidenced by the steeper slope of the regression line.  
In contrast, videos shared in \textit{Politics} groups tend to be shorter, with limited attention given to production quality. Their primary function is to serve immediate needs, such as delivering brief announcements, political speeches, or viral clips intended to shape opinions. A sample of these videos reveals content related to the Russian-Ukrainian conflict, speeches by politicians, and clips with political undertones designed to provoke indignation.

\textit{\textbf{Main takeaways:} video quality (Figure~\ref{fig:video-size-duration}, left) and duration (Figure~\ref{fig:video-size-duration}, right) vary based on the goal of sharing such a video, from entire movies to a few seconds clips to shape the debate. }


\begin{figure}[t]
    \centering
    \includegraphics[width=0.9\columnwidth]{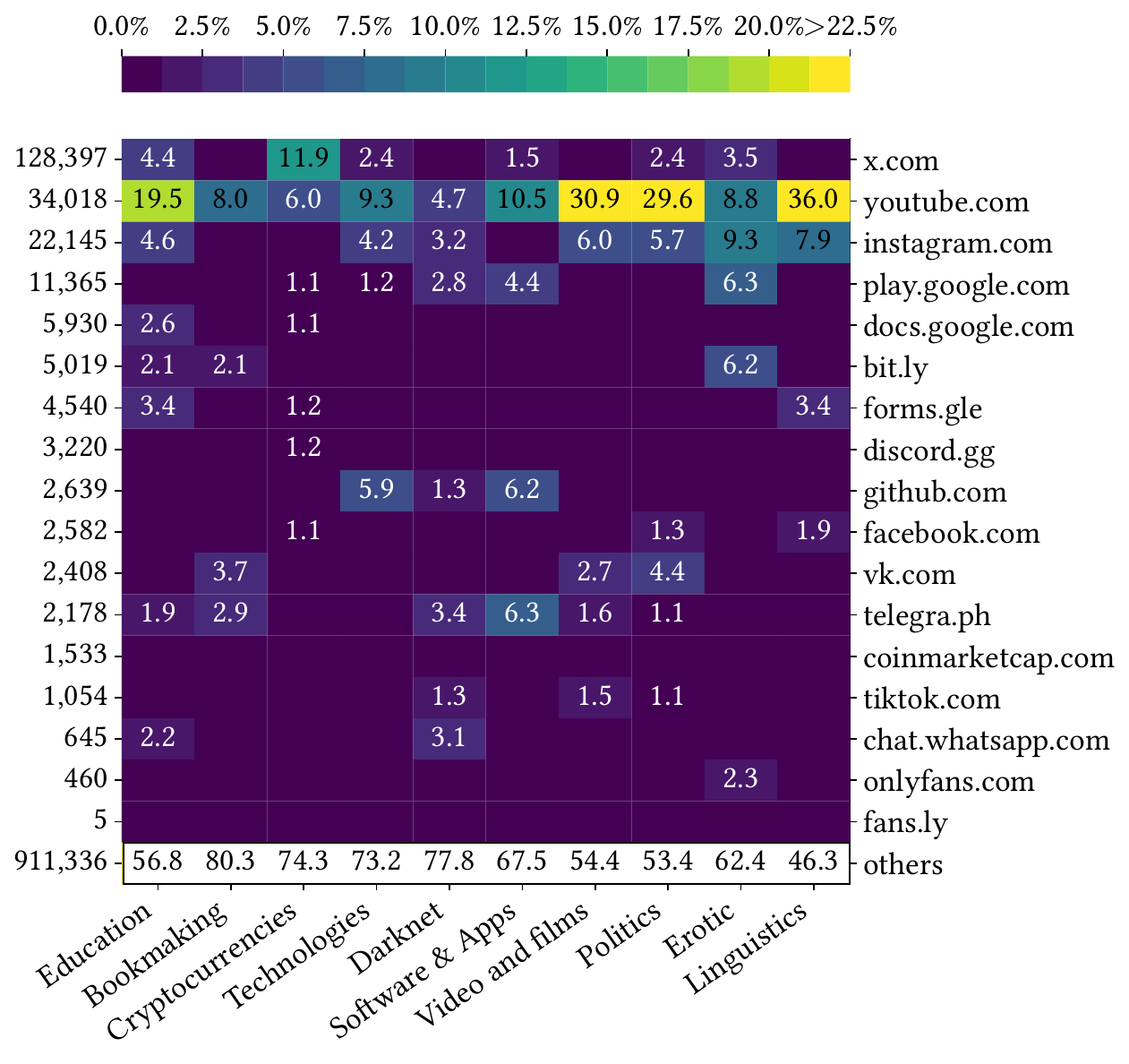}
    \caption{Per-topic average of the percentage of linked domains within groups. On the left, the number of times a domain appears.}
    \label{fig:domain-heatmap}
\end{figure}



\subsection{
Links to external domains}

We now extend the analysis to the sharing of links to external websites (i.e., different from \url{telegram.me} and \url{t.me}) to assess 
how Telegram users wish to redirect (or drive) attention to other websites, and the interplay with external Web resources
.


In Figure~\ref{fig:domain-heatmap}, we present the average frequency at which a domain appears in a given topic. We focus on the union of the 5 most popular domains across each topic. These cover from $\approx$20\% to $\approx$50\% of links --- with a heavy-tailed presence of other platforms (bottom row).
The three most frequent platforms are social networks: X, YouTube and Instagram. Usage varies a lot: X sees significant use in \textit{Cryptocurrency} groups; YouTube and, to a lesser extent, Instagram are the two most transversal platforms. Conversely, in line with their discussion topics, \textit{Technology} and \textit{Software and Application} groups show a significant usage of GitHub pages. The \url{telegra.ph}, an open and anonymous publishing platform, is very often used in \textit{Software \& Applications} groups, e.g., to share anonymously installation guides and tutorials. All in all, we observe a very diverse sharing of information from extensive platforms, with each topic having different preferential means to refer to external outlets.

Finally, we expand \texttt{bit.ly} URLs, finding that only 4.2\% pointed to popular domains (i.e., those in the union of the five most frequent domains per topic), while 67.4\% led to less common websites with few occurrences. Additionally, 28.4\% of these URLs directed to expired domains. As \texttt{bit.ly} URLs represent a small fraction of the total, their expansion minimally impacts domain rankings but rather highlights the use of URL shorteners as a noteworthy behavior.

\textit{\textbf{Main takeaways:} YouTube consistently serves as the most prevalent domain across all topics to share external content (largest share of links across each but one topic). A notable exception is the cryptocurrency discussions, where X takes precedence (11.9\% vs. 6\%). Some platforms are popular in some topics, with again few common habits (except YouTube, see Figure~\ref{fig:domain-heatmap}).}


\vspace{-5pt}
\subsection{Repeated sharing behavior}
\label{subsec:repeated_sharing}

\begin{figure}[t]
    \centering
    \includegraphics[width=0.8\linewidth]{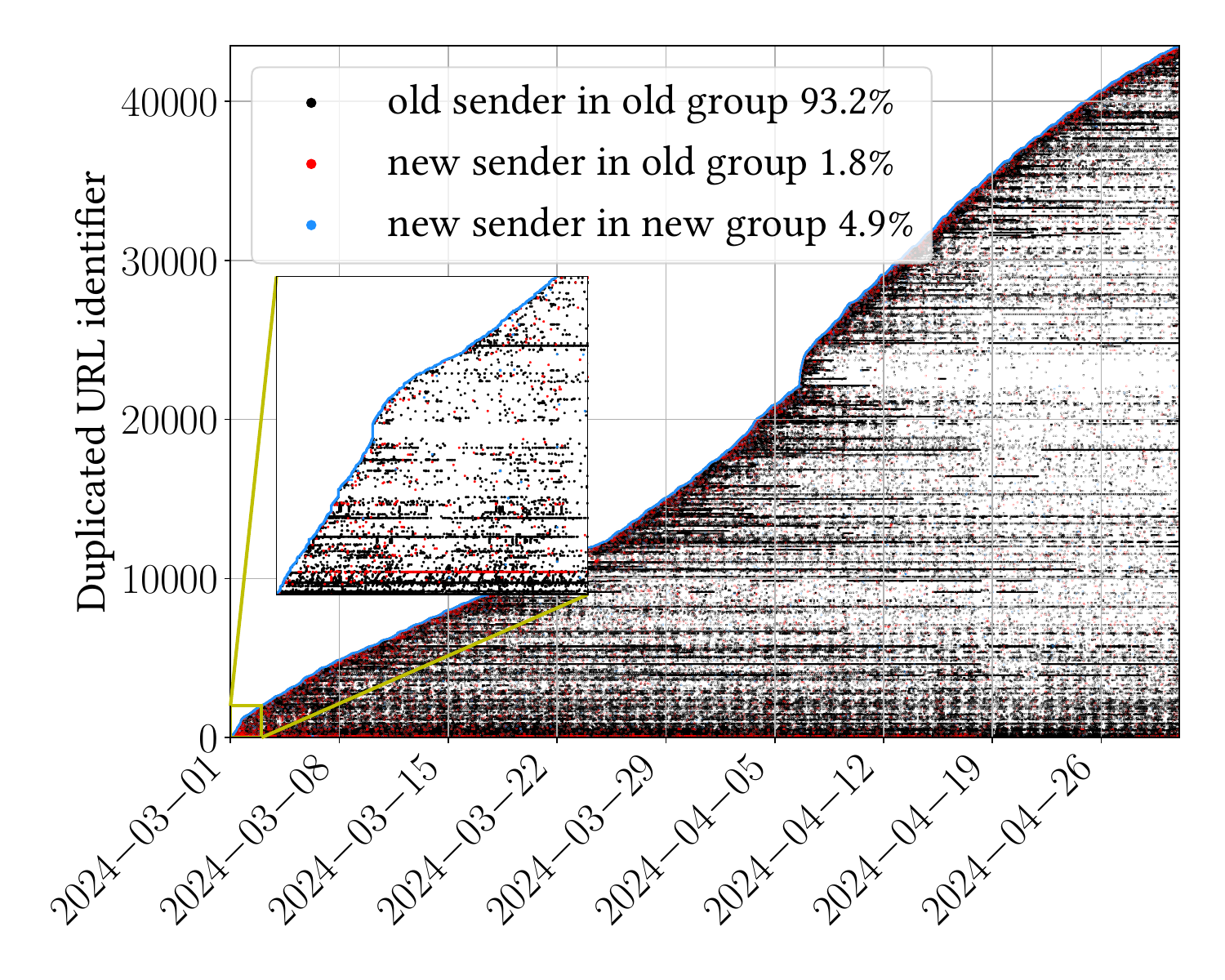}
    \caption{Cumulative time series of unique URLs appearing at least twice in the dataset, along with all their subsequent repetitions. The zoom highlights the first two days. }
    \label{fig:duplicated-url}
\end{figure}

Delving into the analysis of the links to external domains, we now look at possible patterns 
from the repeated sharing of the 
same URL. 
In total, there are  around 239,000 unique URLs  in our dataset, around 43,000 (18\%) of which are shared multiple times during the period of analysis. We here focus on these duplicated  URLs.

In Figure~\ref{fig:duplicated-url} we report a scatter plot where every dot represents the sharing of a given URL (y-axis) at a specific timestamp (x-axis). We sort the unique URLs by their first appearances in our data. 
We distinguish if the URL is shared multiple times by the same sender (\textit{old sender}) or a new one (\textit{new sender}), and if the URL is shared multiple times in the same group (\textit{old group}) or in a new one (\textit{new group}). A sender is \textit{new} when they first send a given URL to a given group. If they send the same URL to the same group again, they become \textit{old}. If they send the same URL to another group for the first time, they are \textit{new} again. 

We find that the large majority (93.2\%) of all duplicated URLs correspond to users posting the same URLs in groups they have already posted the URLs in. This hints at a spamming activity by regular accounts that programmatically share the same URL with the same audience, to maximise the reach of the content. 
Most of these  repeated URLs appear in groups about \textit{Bookmaking} (54\%), \textit{Cryptocurrencies} (21\%) and \textit{Darknet} (14\%) topics. These groups are in fact full of ads and spam messages. 

By manually analysing the most replicated URLs, we observe that many of them advertise Social Media Managing services, which promise to boost one's interactions (followers, likes,
etc.) through the use of likely fake accounts. Curiously, the most repeated URL  appears more than 27,000 times in a single group over two months. 

Our analysis reveals also suspicious patterns. The red line in the inset of Figure~\ref{fig:duplicated-url} shows an uninterrupted sequence of messages with the same URL sent within the same \textit{Bookmaking} group by 1,498 distinct users. This suggests a coordinated network of likely fake users promoting a betting website.
In the first week of April, we also observe a sudden burst of shared URLs that are shared for a few days. These URLs refer to Reddit comments or posts arguably sent by their author who asks for ``upvotes" to other members of a Cryptocurrencies group, suggesting an attempt to manipulate content curation algorithms as shown in \cite{weerasinghe_pod_2020}. 


\begin{figure}[tt!]
    \centering
\includegraphics[width=0.8\linewidth]{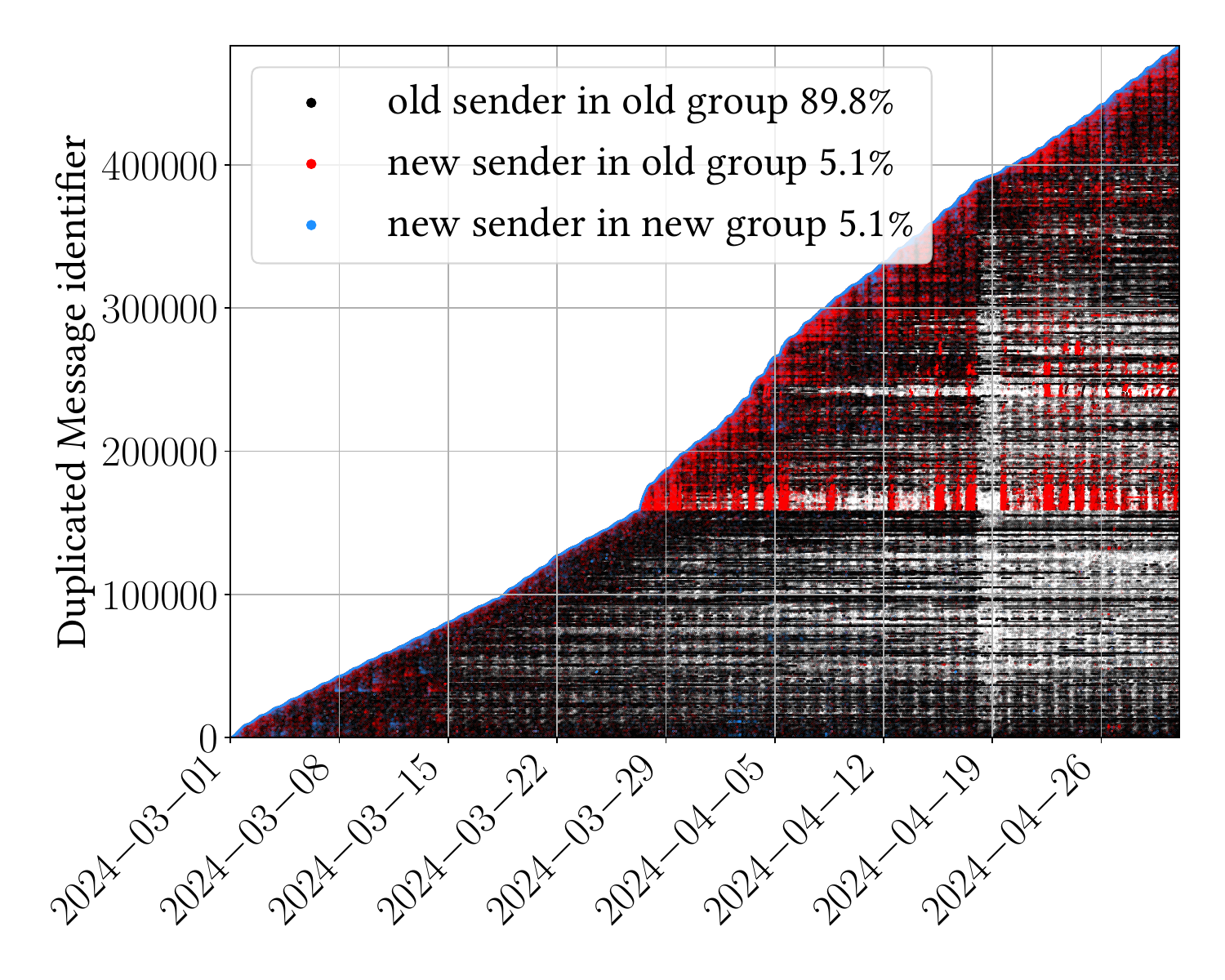}
    \caption{Cumulative time series of unique long messages (over 50 characters) repeated at least twice in the dataset, including all subsequent identical reposts. }
    \label{fig:duplicated-messages}   
\end{figure}


After detecting spamming behaviour in URL sharing, we investigate whether similar patterns occur with textual messages. To filter out trivial repetitions (e.g., "ok" or "good morning"), we focuse on messages longer than 50 characters and analyse perfect duplicates with identical content. As shown in Figure~\ref{fig:duplicated-messages}, duplication is common even for long messages, with nearly 500,000 messages repeated at least once across 97.8\% of the groups, accounting for 65\% of all long messages.
Unlike URLs, duplicated messages are more evenly spread across topics, with the \textit{Darknet} topic showing the highest repetition (19\%) and \textit{Linguistics}, \textit{Video and Films}, and \textit{Education} having the fewest (under 1\%). 

Although Telegram allows message forwarding, only 13.7\% of these repeated messages are forwarded; most are posted directly by the same users. Similar to URL patterns, most duplicated messages come from users repeatedly posting the same content in the same group, further indicating spamming behavior.

Suddenly, starting from the last week of March 2024, a great portion of duplicated messages is reposted in the same group by users who had not previously shared them (\textit{new senders old group}, red pattern).
Interestingly, this phenomenon happens in a single group.
This group (``\texttt{kadyrov\_95chat}'') contributes to 4.3\% of all the repeated messages and 84\% of all the repeated messages sent by different users in the same group. Predominantly in Chechen, messages often include congratulations, generic wishes, and praise for the Chechen Republic and its government. 
We conjecture that this behaviour is caused by some sort of spam bot networks, which tend to copy-paste the same message multiple times. 
However, we acknowledge that we could not retrieve messages sent before March 27\textsuperscript{th}, leaving it unclear whether the group was previously inactive or if earlier messages were deleted before our data collection began. This highlights the importance of tracking a group activity over time, to retrieve its historical messages.


\textit{\textbf{Main takeaways:} we observe indications of spamming behaviour in the sharing of both URLs (Figure~\ref{fig:duplicated-url}) and textual messages (Figure~\ref{fig:duplicated-messages}). Most repeated content comes from the same single users (93.2\% and 89.8\% respectively). Yet, evidence of coordinated group of spammers clearly emerges (e.g., red line in the inset of Figure~\ref{fig:duplicated-url}).
This should encourage further research to characterise this behaviour and its impact on user participation and group dynamics.}

\vspace{-3pt}
\subsection{
Time Lag in YouTube Video Sharing}
\label{fig:YouTube-to-Dataset}

\begin{figure}
    \centering
    \includegraphics[width=0.60\columnwidth]{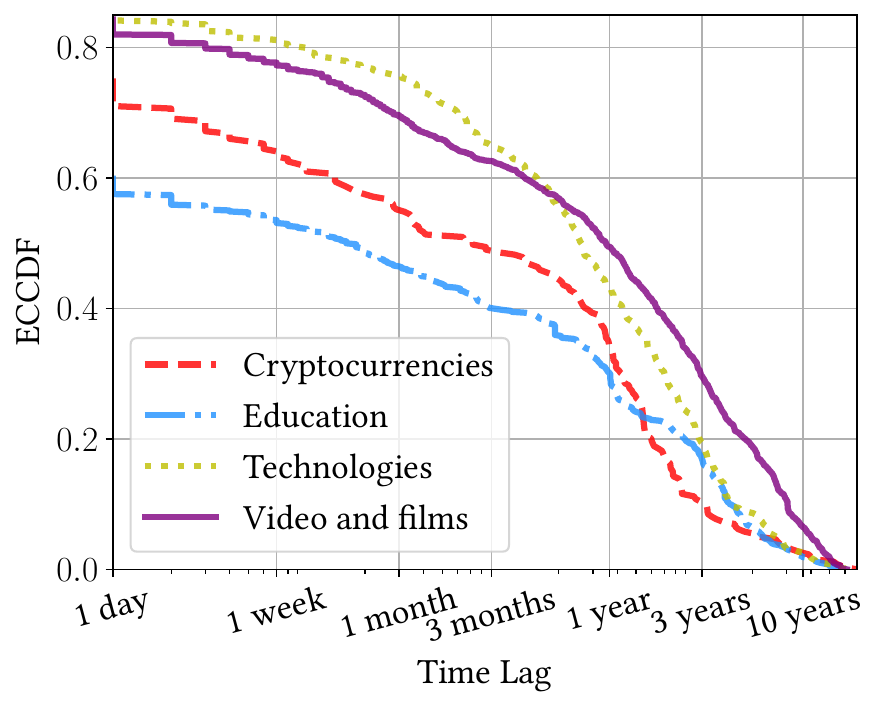}
    \caption{Elapsed time between a video publication on YouTube and its first appearance in Telegram groups, considering 4 topics (log x-scale).}
    \label{fig:YT_delta_T}
\end{figure}

As shown in Figure~\ref{fig:domain-heatmap}, \url{youtube.com} is the most consistent domain shared on Telegram groups across topics. It accounts for the 8.1\% of the unique links shared in our dataset.
For each posted video link, we collect its publication timestamp on YouTube through the \textit{YouTube Data API v3}.\footnote{\url{https://developers.google.com/youtube/v3?hl=en}} Then, we measure the elapsed time to its first appearance in all of our Telegram groups.\footnote{If appearing in multiple groups, we  evaluate the difference multiple times.}  
In Figure~\ref{fig:YT_delta_T}, we present the ECCDF for the topics with the most diverse distributions.
Considering videos shared in {\it Education} groups, more than 40\% of them are shared on Telegram less than 24 hours after their publication on YouTube. This suggests 
a community interested in fresh content. 
Most videos consist of classes and tutorials offering guidance on how to approach specific exams.
The {\it Video and films} topic, in turn,  exhibits a great interest in old videos:
more than 30\% of the videos have been on YouTube for almost three years (mostly old movies). This indicates that this topic is not recency-bound.

In the \textit{Technologies} groups, we observe a limited interest in new videos, with only 20\% of shared content appearing within 10 days of publication. This is surprising, as one would expect a high level of engagement from the community regarding the latest developments in the field. 
One explanation 
lies in the predominance of video tutorials on technology topics, which retain their relevance over time. In fact, most URLs refer to educational content or guides in the technology field, while some present consumer technology reviews.

Finally, it is noteworthy to focus on the knee of the ECCDF of videos shared in the Cryptocurrencies groups: here videos older than one year are less and less shared. This can be due to the variety of Cryptocurrencies and market-related suggestion videos, whose content naturally loses interest over time. 

\textit{\textbf{Main takeaways:} the distribution of the elapsed time between publication on YouTube and sharing on Telegram (Figure~\ref{fig:YT_delta_T}) again showcases context-based patterns that depend on the communities interacting on the platform.}

\section{Conclusion}
\label{sec:conclusion}



In this paper, we conducted a comprehensive, topic-wise analysis of Telegram public groups retrieved from TGStat, providing a transversal view of platform usage across diverse discussion topics. Our work contributes to the understanding of how Telegram is used across various domains, shedding light on usage patterns that had not been explored in prior studies. Furthermore, we develop an open-source crawler and dataset that contributes valuable tools for future researchers to explore Telegram usage.

Our analysis reveals a significant heterogeneity in how different communities leverage the platform's capabilities. Bots, for instance, play a crucial role in many groups, with some topics — such as \textit{Linguistics} and \textit{Education}—showing up to 80–90\% of messages generated by bots. This highlights the importance of automated interaction in certain topics, where bots are integral to group functionality. Conversely, in \textit{Politics} groups, bot activity is notably lower, suggesting higher levels of direct user engagement in discussions.

Media sharing varies significantly across topics. In \textit{Politics} groups, videos are short and focused on quick messages or viral clips, while in \textit{Video and Films} groups, entire movies and TV shows are shared, contributing to a larger data volume. \textit{Erotic} groups prioritize high-resolution videos despite shorter durations. Non-textual elements like stickers, emojis, and links also differ, reflecting diverse community dynamics and interaction styles.

In terms of user activity, the length of user-generated messages differs greatly. In \textit{Darknet} groups, messages are particularly long, often containing detailed information about illicit goods and services, while in discussion-driven groups like \textit{Politics} and \textit{Linguistics} more concise messages prevail. Moreover, we identified substantial repeated sharing behaviour, particularly in \textit{Bookmaking}, \textit{Cryptocurrency}, and \textit{Darknet} groups, where spamming and coordinated actions (e.g., repeated posting of URLs or messages by different users) are common. We highlighted the presence of bot-like networks or coordinated manipulation efforts within specific communities.

Overall, our study offers a broad, multi-faceted exploration of Telegram group usage that contrasts sharply with previous studies, which were more limited in scope. The diversity of patterns we uncovered demonstrates how versatile and complex Telegram usage can be, depending on the topic at hand.

Future research directions 
include identifying and quantifying the key factors that drive user engagement and examining the processes of influence and information dissemination. Additionally, further analysis of content through NLP techniques could shed light on shifts in discussion subjects.

\section{Ethical aspects}
\label{sec:ethics}

In our work, we take ethics under utmost consideration.

$\bullet$  Our data extraction rate on TGStat is below 5 pages per minute, to minimise the load, and we repeat the crawling only once a week.

$\bullet$  We contacted Telegram's support 
    to declare our intention, asking them to share with us any limitation. We received no answer. The privacy policy of the platform does not forbid crawling.

$\bullet$ To respect privacy restrictions imposed by group administrators, we restrict our analysis to public groups.
We do not monitor 
groups where the administrator sets the auto-delete functionality and those where admins refuse or ignore our join request. 
 
 $\bullet$ Telegram might be used to share copyright-protected material and illicit content. We avoid storing any non-textual element, only collecting metadata (e.g., video duration and size).

\section*{Acknowledgements}
This work has been partially supported by the Spoke 1 "FutureHPC \& BigData" of ICSC --- Centro Nazionale di Ricerca in High-Performance-Computing, Big Data and Quantum Computing, funded by European Union --- NextGenerationEU. It has also been partially supported by {\it Conselho Nacional de Desenvolvimento Científico e Tecnológico }(CNPq) and {\it Fundação de Amparo à Pesquisa do Estado de Minas Gerais } (FAPEMIG), both in Brazil.

\bibliographystyle{ACM-Reference-Format}
\bibliography{refs}

\appendix

\section{Appendix}
\label{sec:app}

\begin{table*}[t!]
\footnotesize
\caption{Empirical topic description. TGStat does not provide any description of the topics.}
\begin{tabular}{l|p{13cm}}
\toprule
\textbf{Topics} & \textbf{Description} \\ \hline
\textit{Education} & Discussion about college and university courses and exams. \\ \hline
\textit{Bookmaking} & Discussion about online betting and similar topics. \\ \hline
\textit{Cryptocurrencies} & Discussion about cryptocurrencies, market stock and similar topics. Some groups offer official support for crypto-exchanges. \\ \hline
\textit{Technologies} & Discussions about consumer electronics, mostly smartphones. Some groups are second-hand marketplaces for consumer electronics. \\ \hline
\textit{Darknet} & Trading of mostly illegal content, i.e., credit card numbers, accounts of media platforms, etc. \\ \hline
\textit{Software and apps} & Discussion about usage of software, mostly regarding Android modding and app piracy. Some groups discuss software development for specific languages or technologies.  \\ \hline
\textit{Video and Films} & Discussion about movies and video sharing of movies and TV series and (possibly piracy). \\ \hline
\textit{Politics} & Discussion about political news at large in different countries. \\ \hline
\textit{Erotic} & Sharing and suggestion of adult content or services.  \\ \hline
\textit{Linguistics} & Community of users practising a particular language for educational purposes. \\ \hline
\textit{Courses and guides} & Some groups share content related to MOOCs or paid online courses, but most groups are filled with spam advertising new courses.  \\ \hline
\textit{Economics} & Discussion about market stocks and investments, but most groups are actually about cryptocurrencies. \\
\bottomrule
\end{tabular}
\label{tab:topic-description}
\end{table*}

Appendix includes additional results to contextualize our analysis. 

\begin{figure}[b!]
    \centering
    \includegraphics[width=0.7\linewidth]{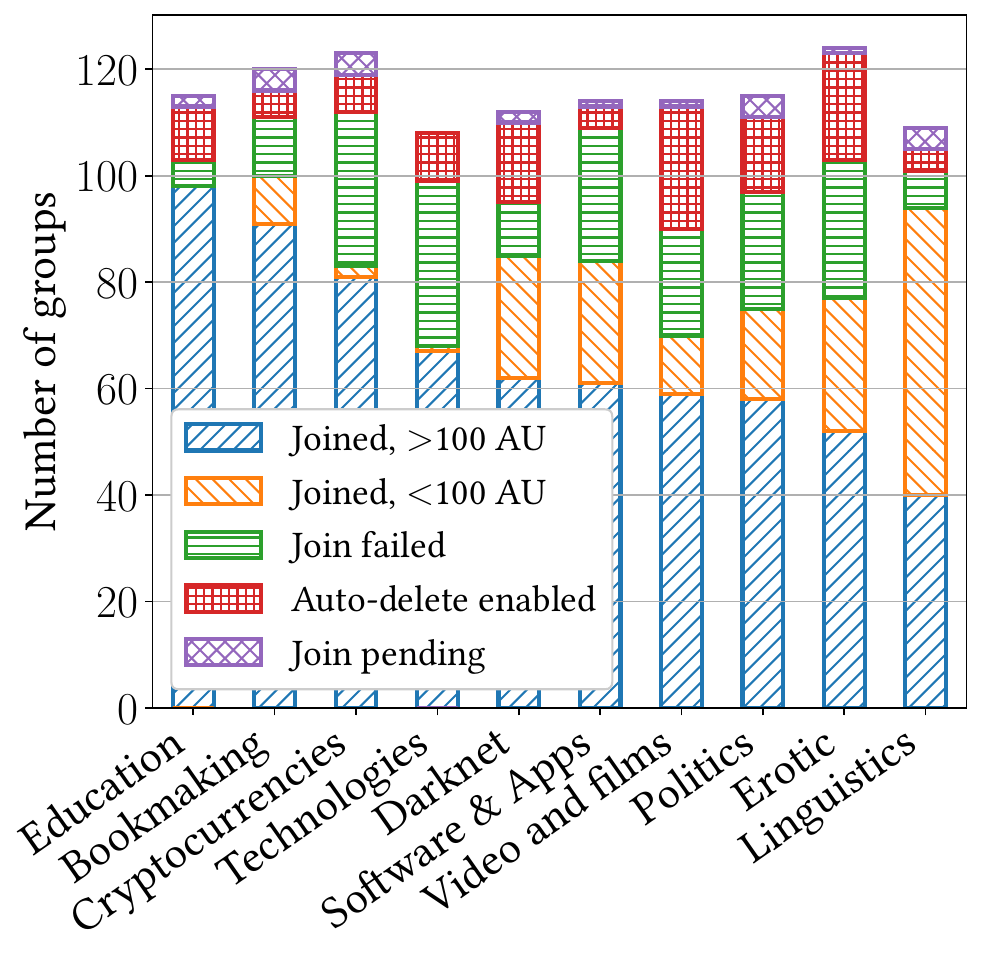}
    \caption{Per-topic groups of TGStat and results of our crawling.}
    \label{fig:crawler_result}
\end{figure}

In Figure~\ref{fig:crawler_result}, we show the breakdown of the groups we found on TGStat, and those we actually monitor for the paper (blue pattern). The figure details the fraction of groups we ignored for various motivations. 

In Table~\ref{tab:topic-description}, we provide a qualitative explanation of the topics under observation. In 

Table~\ref{tab:details_more}, we present additional per-topic detailed statistics.

\begin{figure}[b!]
    \centering
    \begin{subfigure}{.4\columnwidth}
        \includegraphics[width=\textwidth]{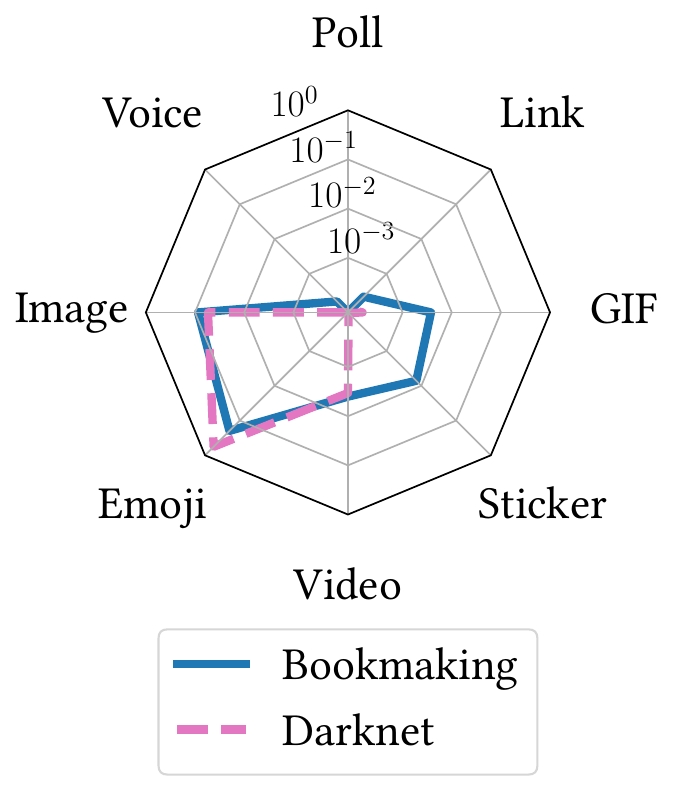}
        \label{fig:spiderplot-bookmaking-darknet}
    \end{subfigure}
    \begin{subfigure}{.4\columnwidth}
    \includegraphics[width=\textwidth]{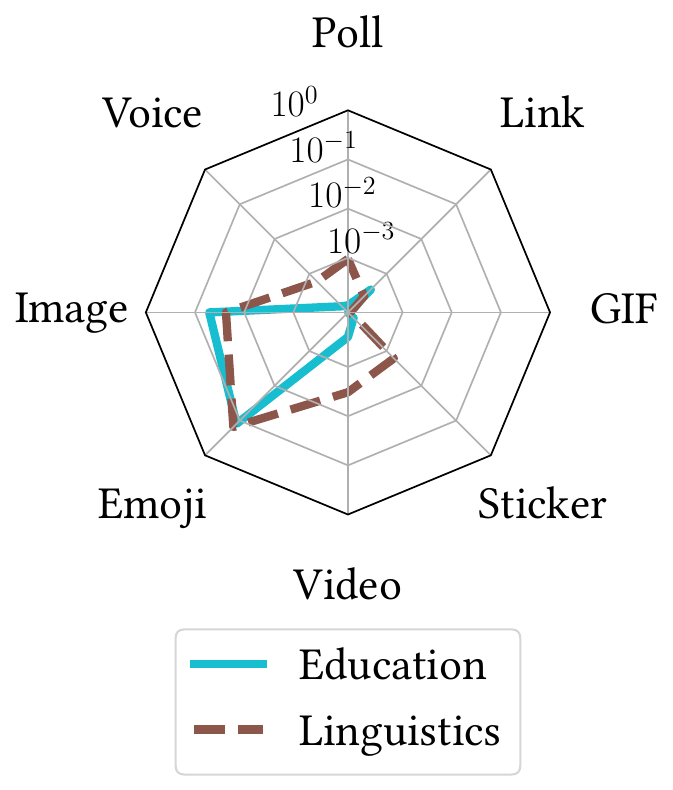}
        \label{fig:spiderplot-education-linguistics}
    \end{subfigure}
    \\
    \begin{subfigure}{.4\columnwidth}
        \includegraphics[width=\textwidth]{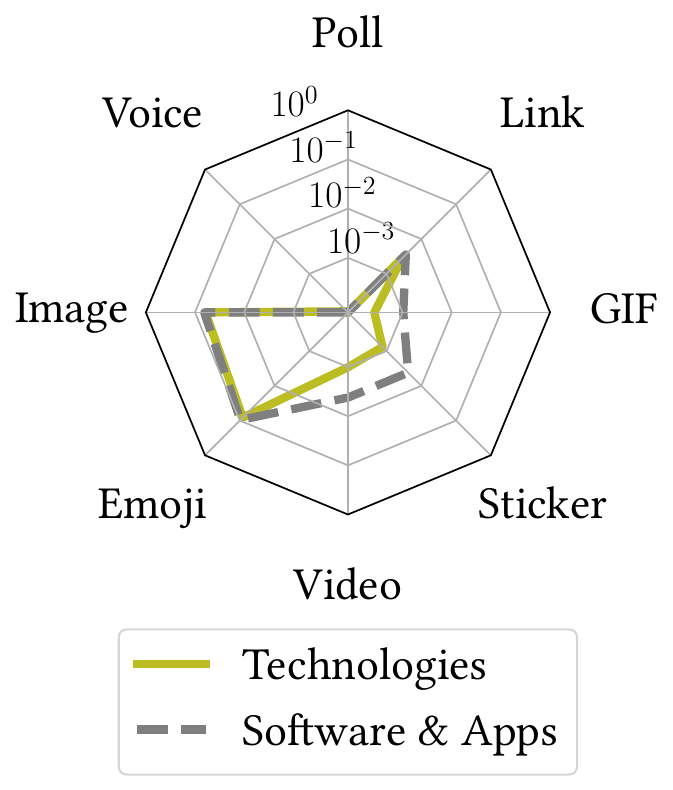}
        \label{fig:spiderplot-technologies-software-and-apps}
    \end{subfigure}
    \caption{Median fraction of messages with non-textual elements in topics left out from Figure~\ref{fig:media-spiderplots}.
    }
    \label{fig:appendix-spiderplots}
\end{figure}

In Figure~\ref{fig:appendix-spiderplots}, we compare the median use of non-textual media in group messages by topic.

\begin{table}[b!]
\caption{Per-topic additional metrics.}
\footnotesize
\begin{tabular}{l|rr|r|r|cc}
\toprule
\textbf{Topic} & 
\multicolumn{1}{p{0.8cm}}{\centering \textbf{English groups}} & 
\multicolumn{1}{p{0.8cm}|}{\centering \textbf{Russian groups}} & 
\multicolumn{1}{p{.6cm}|}{\centering \textbf{Bot msgs (\%)}} & 
\multicolumn{1}{p{.9cm}|}{\centering \textbf{Avg msg length (ch.)}} & 
\multicolumn{1}{p{.8cm}}{\centering \textbf{Avg video size (MB)}} & 
\multicolumn{1}{p{1cm}}{\centering \textbf{Avg video duration (min)}} \\ \hline
\textit{Education} & 46 & 13 & 8.9 & 43.3  & 49.9 & 4.3 \\
\textit{Bookmaking} & 24 & 49 & 6.4 & 68.3 &  14.1 & 0.8 \\
\textit{Crypto} & 45 & 15 & 8.3 & 44.4 & 12.5 & 0.9 \\
\textit{Technologies} & 23 & 30 & 9.9  & 58.5 & 13.1 & 0.8 \\
\textit{Darknet} & 9 & 40 & 8.0 & 305.4 & 7.8 & 0.5 \\
\textit{Software} & 14 & 28 & 5.4 & 40.1 & 17.4 & 1.4 \\
\textit{Video\&Films} & 21 & 23 & 9.6 & 61.1 &  48.6 & 5.1 \\
\textit{Politics} & 11 & 34 & 3.6 & 160.8 & 21.2 & 1.9 \\
\textit{Erotic} & 15 & 14 & 6.6 & 60.1 & 60.1 & 2.8 \\
\textit{Linguistics} & 20 & 11 & 12.6 & 35.0 & 19.5 & 3.4 \\ \hline
\textit{All Topics} & 228 & 257 & 7.8 & 65.5 & 26.1 & 2.1 \\ \bottomrule
\end{tabular}
\label{tab:details_more}
\end{table}

\end{document}